\documentclass[12pt,executivepaper]{article}
\usepackage{amssymb}
\usepackage{amsmath}

\usepackage{aeguill}

\input{tcilatex}

\begin{document}

\begin{center}
{\Large The effective potential and resummation procedure to
multidimensional complex cubic potentials for weak and strong-coupling}

\bigskip

S.-A. Yahiaoui, O. Cherroud, M. Bentaiba\footnote{%
Corresponding author
\par
E-mail address : bentaiba@hotmail.com}

LPTHIRM, D\'{e}partement de Physique, Facult\'{e} des Sciences,

Universit\'{e} Saad DAHLAB de blida, Alg\'{e}rie.

\bigskip

\textbf{Abstract}
\end{center}

The method for the recursive calculation of the effective potential is
applied successfully in case of weak coupling limit (g tend to zero) to a
multidimensional complex cubic potential. In strong-coupling limit (g tend
to infinity), the result is resumed using the variational perturbation
theory (VPT). It is found that the convergence of VPT-results approaches
those expected.

\bigskip

PACS: 03.65.Ca; 05.30.-d; 03.65.-w; 02.30.Mv.

KeyWords: Variational perturbation theory; Effective potential; Feynman
diagrams.

\bigskip

\section{Introduction}

Quantum dynamics allows to avoid the operator formalism for the benefit of
infinite products of integrals, called Path-Integrals [1,2]. Unfortunately,
most Path-Integrals can not be performed exactly; therefore, many different
approximation procedures are developed in order to deal with
non-analytically solvable systems.

The most commonly used one is known as perturbation theory [3-9]. It is
based upon the expansion of some physical quantity, e.g. the ground-state
energy of particle in some potential, into a power series of coupling
constant, the results obtained in weak-coupling limit seem to converge to
the exact result for low orders, so that the divergence of that series
becomes important when the expansion is driven to the higher-order [3-12].
Therefore, it is necessary to find means for treating the divergent
perturbation series, including the strong-coupling limit. Thus, we apply the
resummation scheme often based on introducing an artificial parameter trick,
namely Kleinert's square-root [2,13-15].

The variational perturbation theory (VPT) [2,7-9] has been developed thanks
to the variational approach due to R. P. Feynman and H. Kleinert [6], and
has been extended to an efficient non-perturbation approximation. VPT allows
the conversion of the divergent weak-coupling into the convergent
strong-coupling expansion.

According to some studies, a more physical alternative axiom called
space-time reflection symmetry $\left( \mathcal{PT-}\text{symmetry}\right) $
allows for possibility of cubic complex non-Hermitian Hamiltonian but still
leads to a consistant theory of quantum mechanics. It is shown that if the $%
\mathcal{PT-}$symmetry of a Hamiltonian $H$ is not broken, then the
accompanying spectrum remains real and positive.

The space reflection operator $\mathcal{P}$ is a linear operator with the
property $\mathcal{P}^{2}=1$ and has the effects $p\rightarrow -p$, and $%
x\rightarrow -x$, while the time reflection $\mathcal{T}$ is an antilinear
operator with the property $\mathcal{T}^{2}=1$ and has the effects $%
p\rightarrow -p$, $x\rightarrow x$, and $i\rightarrow -i$.

Recently, the convergence of VPT has been tested successfully for the
ground-state energy of the one-dimensional complex cubic coupled oscillator
[11,13,15]%
\begin{equation}
V\left( x\right) =\frac{M}{2}\omega ^{2}x^{2}+igx^{3}.  \tag{1}
\end{equation}%
The purpose of this paper is to examine how, in a first approach, the VPT
can be applied to the generalized 2D and 3D-complex cubic potentials%
\begin{equation}
V\left( x,y\right) =\frac{M}{2}\omega ^{2}\left( x^{2}+y^{2}\right)
+igxy^{2}.  \tag{2}
\end{equation}%
\begin{equation}
V\left( x,y,z\right) =\frac{M}{2}\omega ^{2}\left( x^{2}+y^{2}+z^{2}\right)
+igxyz.  \tag{3}
\end{equation}%
by resuming the weak-coupling series [10,12] of the ground-state energy via
VPT.

\ In the strong-coupling limit, the potentials (2) and (3) are reduced to
potentials without a harmonic term. It turns out that the rate of
convergence is not satisfactory. Therefore, combining the effective
potential [13-15] to the VPT permits the improvement of the rate of
convergence. This later should be performed to higher orders, and since
mathematical computation become cumbersome, we have stopped our calculation
at the second order in $\hbar $.

\section{Perturbation theory : Feynman diagrams}

In this section, we derive the weak-coupling coefficients for the
ground-state energy of the potentials (2) and (3) using the Feynman
diagrammatical expansion. The partition functions%
\begin{equation}
Z_{2D}=\oint \mathcal{D}x\mathcal{D}y\exp \left\{ -\frac{1}{\hbar }%
\int\limits_{0}^{\hbar \beta }d\tau \left[ \frac{M}{2}\left( \overset{\cdot }%
{x}^{2}+\overset{\cdot }{y}^{2}\right) +\frac{M\omega ^{2}}{2}\left(
x^{2}+y^{2}\right) +igxy^{2}\right] \right\} ,  \tag{4}
\end{equation}%
\begin{gather}
Z_{3D}=\oint \mathcal{D}x\mathcal{D}y\mathcal{D}z\exp \left\{ -\frac{1}{%
\hbar }\int\limits_{0}^{\hbar \beta }d\tau \left[ \frac{M}{2}\left( \overset{%
\cdot }{x}^{2}+\overset{\cdot }{y}^{2}+\overset{\cdot }{z}^{2}\right)
\right. \right.  \notag \\
\left. \left. +\frac{M\omega ^{2}}{2}\left( x^{2}+y^{2}+z^{2}\right) +igxyz%
\right] \right\} .  \tag{5}
\end{gather}%
will be calculated perturbatively by an expansion in the coupling constant $%
g $, here we have used $x_{\alpha }$ (resp. $\overset{\cdot }{x}_{\alpha }$)
instead of $x_{\alpha }\left( \tau \right) $ (resp. $\overset{\cdot }{x}%
_{\alpha }\left( \tau \right) $\ ), and dot $^{\text{\guillemotleft }\ \cdot
\ \text{\guillemotright }}$ refers to derivatives of $x_{\alpha }\left( \tau
\right) $ with respect to $\tau $. These expressions can be converted, using
the Taylor expansion of the exponential function, into the form

\begin{gather}
Z_{2D}=\oint \mathcal{D}x\mathcal{D}y\exp \left\{ -\frac{1}{\hbar }%
\int\limits_{0}^{\hbar \beta }d\tau \left[ \frac{M}{2}\left( \overset{\cdot }%
{x}^{2}+\overset{\cdot }{y}^{2}\right) +\frac{M\omega ^{2}}{2}\left(
x^{2}+y^{2}\right) \right] \right\}  \notag \\
\times \left( 1-\frac{1}{\hbar }\int\limits_{0}^{\hbar \beta }d\tau _{1}%
\left[ igx\left( \tau _{1}\right) y^{2}\left( \tau _{1}\right) \right]
\right.  \notag \\
\left. +\frac{1}{2\hbar ^{2}}\int\limits_{0}^{\hbar \beta }d\tau
_{1}\int\limits_{0}^{\hbar \beta }d\tau _{2}\left[ igx\left( \tau
_{1}\right) y^{2}\left( \tau _{1}\right) \right] \left[ igx\left( \tau
_{2}\right) y^{2}\left( \tau _{2}\right) \right] +\cdots \right) ,  \tag{6}
\end{gather}

\begin{gather}
Z_{3D}=\oint \mathcal{D}x\mathcal{D}y\mathcal{D}z\exp \left\{ -\frac{1}{%
\hbar }\int\limits_{0}^{\hbar \beta }d\tau \left[ \frac{M}{2}\left( \overset{%
\cdot }{x}^{2}+\overset{\cdot }{y}^{2}+\overset{\cdot }{z}^{2}\right) +\frac{%
M\omega ^{2}}{2}\left( x^{2}+y^{2}+z^{2}\right) \right] \right\}  \notag \\
\times \left( 1-\frac{1}{\hbar }\int\limits_{0}^{\hbar \beta }d\tau _{1}%
\left[ igx\left( \tau _{1}\right) y\left( \tau _{1}\right) z\left( \tau
_{1}\right) \right] +\right.  \notag \\
\left. \frac{1}{2\hbar ^{2}}\int\limits_{0}^{\hbar \beta }d\tau
_{1}\int\limits_{0}^{\hbar \beta }d\tau _{2}\left[ igx\left( \tau
_{1}\right) y\left( \tau _{1}\right) z\left( \tau _{1}\right) \right] \left[
igx\left( \tau _{2}\right) y\left( \tau _{2}\right) z\left( \tau _{2}\right) %
\right] +\cdots \right) .  \tag{7}
\end{gather}

Introducing in both cases the notation, namely expectation values,%
\begin{equation}
\left\langle \cdots \right\rangle _{\omega }\equiv \frac{1}{%
\prod\limits_{\alpha =1}^{3}Z_{x_{\alpha }}^{\omega }}\oint \prod_{\alpha
=1}^{3}\mathcal{D}x_{\alpha }\left( \cdots \right) \exp \left\{ -\frac{1}{%
\hbar }\int\limits_{0}^{\hbar \beta }d\tau \sum_{\alpha =1}^{3}\left( \frac{M%
}{2}\overset{\cdot }{x}_{\alpha }^{2}+\frac{M}{2}x_{\alpha }^{2}\right)
\right\} ,  \tag{8}
\end{equation}%
lead, up to the second order in $g,$\ to the following expressions of the
partition functions%
\begin{equation}
Z_{2D}\approx Z_{x}^{\omega }Z_{y}^{\omega }\left[ 1-\frac{g^{2}}{2\hbar ^{2}%
}\int\limits_{0}^{\hbar \beta }d\tau _{1}\int\limits_{0}^{\hbar \beta }d\tau
_{2}\left\langle x\left( \tau _{1}\right) y^{2}\left( \tau _{1}\right)
x\left( \tau _{2}\right) y^{2}\left( \tau _{2}\right) \right\rangle _{\omega
}\right] ,  \tag{9}
\end{equation}%
\begin{equation}
Z_{3D}\approx Z_{x}^{\omega }Z_{y}^{\omega }Z_{z}^{\omega }\left[ 1-\frac{%
g^{2}}{2\hbar ^{2}}\int\limits_{0}^{\hbar \beta }d\tau
_{1}\int\limits_{0}^{\hbar \beta }d\tau _{2}\left\langle x\left( \tau
_{1}\right) y\left( \tau _{1}\right) z\left( \tau _{1}\right) x\left( \tau
_{2}\right) y\left( \tau _{2}\right) z\left( \tau _{2}\right) \right\rangle
_{\omega }\right] ,  \tag{10}
\end{equation}%
where $Z_{x_{\alpha }}^{\omega }$ are the partition functions of the
harmonic oscillator%
\begin{equation}
Z_{x}^{\omega }=Z_{y}^{\omega }=Z_{z}^{\omega }=\frac{1}{2}\sinh ^{-1}\frac{%
\hbar \beta \omega }{2}.  \tag{11}
\end{equation}%
Expectation values in (9) and (10) can be performed by applying generalized
Wick's rules

\begin{itemize}
\item each expectation values of a product of $2$-paths defines the
propagator%
\begin{equation}
G_{\omega }\left( \tau _{1},\tau _{2}\right) =\left\{ \QATOP{\left\langle
x_{i}\left( \tau _{1}\right) x_{j}\left( \tau _{2}\right) \right\rangle
_{\omega }\neq 0;\text{ \ \ for\textrm{\ \ \ }}i=j.}{\left\langle
x_{i}\left( \tau _{1}\right) x_{j}\left( \tau _{2}\right) \right\rangle
_{\omega }=0;\text{ \ \ for\textrm{\ \ \ }}i\neq j.}\right.  \tag{12}
\end{equation}

\item each expectation values of a product of $n$-paths reads, taking into
account the first rule (12)%
\begin{align}
\left\langle x_{i}\left( \tau _{1}\right) x_{j}\left( \tau _{2}\right)
\cdots x_{q}\left( \tau _{n}\right) \right\rangle _{\omega }& =G_{\omega
}\left( \tau _{1},\tau _{2}\right) \left\langle x_{k}\left( \tau _{3}\right)
x_{l}\left( \tau _{4}\right) \cdots x_{q}\left( \tau _{n}\right)
\right\rangle _{\omega }  \notag \\
& +G_{\omega }\left( \tau _{1},\tau _{3}\right) \left\langle x_{j}\left(
\tau _{2}\right) \cdots x_{q}\left( \tau _{n}\right) \right\rangle _{\omega
}+\cdots  \notag \\
& +G_{\omega }\left( \tau _{1},\tau _{n}\right) \left\langle x_{j}\left(
\tau _{2}\right) \cdots x_{p}\left( \tau _{n-1}\right) \right\rangle
_{\omega },  \tag{13}
\end{align}
\end{itemize}

By applying the Wick's rules and knowing that the free energy reads as $%
F=-k_{B}T\ \ln Z$, we derive the perturbation series at low temperatures $%
\left( T\rightarrow 0\right) $ for the ground-state energy of the potentials
(2) and (3) up to the fourth-order by the connected Feynman diagrams%
\begin{eqnarray}
E_{2D} &=&\hbar \ \omega -\underset{T\rightarrow 0}{\lim }k_{B}\ T\ \left[ 
\frac{1}{72}\FRAME{itbpF}{1.173cm}{0.5623cm}{0.2109cm}{}{}{f_1.jpg}{\special%
{language "Scientific Word";type "GRAPHIC";maintain-aspect-ratio
TRUE;display "PICT";valid_file "F";width 1.173cm;height 0.5623cm;depth
0.2109cm;original-width 0.4272in;original-height 0.1877in;cropleft
"0";croptop "1.0259";cropright "1.0097";cropbottom "0";filename
'Feyndiag/f_1.JPG';file-properties "XNPEU";}}+\frac{1}{36}\FRAME{itbpF}{%
0.6129cm}{0.6129cm}{0.2109cm}{}{}{f_2.jpg}{\special{language "Scientific
Word";type "GRAPHIC";maintain-aspect-ratio TRUE;display "PICT";valid_file
"F";width 0.6129cm;height 0.6129cm;depth 0.2109cm;original-width
0.2084in;original-height 0.2084in;cropleft "0";croptop "1.0200";cropright
"1.0200";cropbottom "0";filename 'Feyndiag/f_2.JPG';file-properties "XNPEU";}%
}+\frac{1}{324}\FRAME{itbpF}{1.1466cm}{0.6129cm}{0.2109cm}{}{}{f_3.jpg}{%
\special{language "Scientific Word";type "GRAPHIC";maintain-aspect-ratio
TRUE;display "PICT";valid_file "F";width 1.1466cm;height 0.6129cm;depth
0.2109cm;original-width 0.4168in;original-height 0.2084in;cropleft
"0";croptop "1.0200";cropright "1.0116";cropbottom "0";filename
'Feyndiag/f_3.JPG';file-properties "XNPEU";}}+\frac{5}{1296}\FRAME{itbpF}{%
0.5865cm}{0.6129cm}{0.2109cm}{}{}{f_4.jpg}{\special{language "Scientific
Word";type "GRAPHIC";maintain-aspect-ratio TRUE;display "PICT";valid_file
"F";width 0.5865cm;height 0.6129cm;depth 0.2109cm;original-width
0.198in;original-height 0.2084in;cropleft "0";croptop "1.0200";cropright
"1.0210";cropbottom "0";filename 'Feyndiag/f_4.JPG';file-properties "XNPEU";}%
}\right.  \notag \\
&&\left. +\frac{1}{648}\FRAME{itbpF}{0.5865cm}{0.6129cm}{0.2109cm}{}{}{%
f_5.jpg}{\special{language "Scientific Word";type
"GRAPHIC";maintain-aspect-ratio TRUE;display "PICT";valid_file "F";width
0.5865cm;height 0.6129cm;depth 0.2109cm;original-width
0.198in;original-height 0.2084in;cropleft "0";croptop "1.0200";cropright
"1.0210";cropbottom "0";filename 'Feyndiag/f_5.JPG';file-properties "XNPEU";}%
}+\frac{1}{1296}\FRAME{itbpF}{1.709cm}{0.6129cm}{0.2109cm}{}{}{f_6.jpg}{%
\special{language "Scientific Word";type "GRAPHIC";maintain-aspect-ratio
TRUE;display "PICT";valid_file "F";width 1.709cm;height 0.6129cm;depth
0.2109cm;original-width 0.646in;original-height 0.2084in;cropleft
"0";croptop "1.0200";cropright "0.9935";cropbottom "0";filename
'Feyndiag/f_6.JPG';file-properties "XNPEU";}}\right] +O\left( g^{6}\right) ,
\TCItag{14}
\end{eqnarray}%
\begin{equation}
E_{3D}=\frac{3}{2}\ \hbar \ \omega -\underset{T\rightarrow 0}{\lim }k_{B}T\ %
\left[ \frac{1}{72}\FRAME{itbpF}{0.6129cm}{0.6129cm}{0.2109cm}{}{}{f_2.jpg}{%
\special{language "Scientific Word";type "GRAPHIC";maintain-aspect-ratio
TRUE;display "PICT";valid_file "F";width 0.6129cm;height 0.6129cm;depth
0.2109cm;original-width 0.2084in;original-height 0.2084in;cropleft
"0";croptop "1.0200";cropright "1.0200";cropbottom "0";filename
'Feyndiag/f_2.JPG';file-properties "XNPEU";}}+\frac{1}{1728}\FRAME{itbpF}{%
0.5865cm}{0.6129cm}{0.2109cm}{}{}{f_4.jpg}{\special{language "Scientific
Word";type "GRAPHIC";maintain-aspect-ratio TRUE;display "PICT";valid_file
"F";width 0.5865cm;height 0.6129cm;depth 0.2109cm;original-width
0.198in;original-height 0.2084in;cropleft "0";croptop "1.0200";cropright
"1.0210";cropbottom "0";filename 'Feyndiag/f_4.JPG';file-properties "XNPEU";}%
}+\frac{1}{5184}\FRAME{itbpF}{0.5865cm}{0.6129cm}{0.2109cm}{}{}{f_5.jpg}{%
\special{language "Scientific Word";type "GRAPHIC";maintain-aspect-ratio
TRUE;display "PICT";valid_file "F";width 0.5865cm;height 0.6129cm;depth
0.2109cm;original-width 0.198in;original-height 0.2084in;cropleft
"0";croptop "1.0200";cropright "1.0210";cropbottom "0";filename
'Feyndiag/f_5.JPG';file-properties "XNPEU";}}\right] +O\left( g^{6}\right) .
\tag{15}
\end{equation}%
For evaluating the connected Feynman diagrams, the following Feynman rules
are introduced [2,14]:

\begin{itemize}
\item for propagator%
\begin{equation}
\FRAME{itbpF}{2.4229cm}{0.9951cm}{0.5008cm}{}{}{f_7.jpg}{\special{language
"Scientific Word";type "GRAPHIC";maintain-aspect-ratio TRUE;display
"PICT";valid_file "F";width 2.4229cm;height 0.9951cm;depth
0.5008cm;original-width 0.9167in;original-height 0.3649in;cropleft
"0";croptop "0.9942";cropright "1.0045";cropbottom "0";filename
'Feyndiag/f_7.JPG';file-properties "XNPEU";}}\rightarrow \quad G_{\omega
}\left( \tau _{a},\tau _{b}\right) =\frac{\hbar }{2\omega }\exp \left[
-\omega \left\vert \tau _{a}-\tau _{b}\right\vert \right] .  \tag{16}
\end{equation}

\item for vertices%
\begin{equation}
\FRAME{itbpF}{1.8364cm}{1.81cm}{0.905cm}{}{}{f_8.jpg}{\special{language
"Scientific Word";type "GRAPHIC";maintain-aspect-ratio TRUE;display
"PICT";valid_file "F";width 1.8364cm;height 1.81cm;depth
0.905cm;original-width 0.6875in;original-height 0.6772in;cropleft
"0";croptop "1.0061";cropright "1.0060";cropbottom "0";filename
'Feyndiag/f_8.JPG';file-properties "XNPEU";}}\rightarrow \quad \frac{-6\ i\ g%
}{\hbar }\int_{0}^{\hbar \beta }d\tau _{a}.  \tag{17}
\end{equation}
\end{itemize}

Applying the Feynman Rules to (4) and (5), we obtain the analytical
expressions for the ground-state energy [10,12]%
\begin{equation}
\ E_{2D}^{\left( 0\right) }=\hbar \omega +\frac{5}{24}\frac{g^{2}\hbar ^{2}}{%
M^{3}\omega ^{4}}-\frac{223}{864}\frac{g^{4}\hbar ^{3}}{M^{6}\omega ^{9}}%
+O\left( g^{6}\right) .  \tag{18}
\end{equation}%
\begin{equation}
E_{3D}^{\left( 0\right) }=\frac{3}{2}\hbar \omega +\frac{1}{24}\frac{%
g^{2}\hbar ^{2}}{M^{3}\omega ^{4}}-\frac{7}{576}\frac{g^{4}\hbar ^{3}}{%
M^{6}\omega ^{9}}+O\left( g^{6}\right) .  \tag{19}
\end{equation}

\section{Bender-Wu perturbation theory : Recursion relations}

We derive here recursion relations of the perturbation coefficients of the
ground-state energy for the potentials (2) and (3). These recursion
relations are obtained from the corresponding Schr\"{o}dinger equations%
\begin{equation}
-\frac{\hbar ^{2}}{2M}\triangledown _{2D}^{2}\psi \left( x,y\right) +\left[ 
\frac{M\omega ^{2}}{2}\left( x^{2}+y^{2}\right) +igxy^{2}\right] \psi \left(
x,y\right) =E_{2D}\psi \left( x,y\right) .  \tag{20}
\end{equation}%
\begin{equation}
-\frac{\hbar ^{2}}{2M}\triangledown _{3D}^{2}\psi \left( x,y,z\right) +\left[
\frac{M\omega ^{2}}{2}\left( x^{2}+y^{2}+z^{2}\right) +igxyz\right] \psi
\left( x,y,z\right) =E_{3D}\psi \left( x,y,z\right) .  \tag{21}
\end{equation}

In the perturbation theory, the ground-state wave functions are expanded in
the form [3-5,12]%
\begin{equation}
\psi \left( x,y\right) =\mathcal{N}_{0}\exp \left[ -\frac{M\omega }{2\hbar }%
\left( x^{2}+y^{2}\right) +\phi \left( x,y\right) \right] .  \tag{22}
\end{equation}%
\begin{equation}
\psi \left( x,y,z\right) =\mathcal{N}_{0}\exp \left[ -\frac{M\omega }{2\hbar 
}\left( x^{2}+y^{2}+z^{2}\right) +\phi \left( x,y,z\right) \right] . 
\tag{23}
\end{equation}

The wave functions $\phi \left( x_{\alpha }\right) ,$ where $\alpha =1,2,3,$
will be expanded in powers of the coupling constant $g$%
\begin{equation}
\phi \left( x_{\alpha }\right) =\sum_{k=1}^{\infty }g^{k}\phi _{k}\left(
x_{\alpha }\right) .  \tag{24}
\end{equation}

The ground-state energy corresponding to (2) and (3) can be expanded in
powers of the coupling constant $g$%
\begin{equation}
E_{2D}=\hbar \omega +\sum_{k=1}^{\infty }g^{k}\epsilon _{k}^{\left(
2D\right) }.  \tag{25}
\end{equation}%
\begin{equation}
E_{3D}=\frac{3}{2}\hbar \omega +\sum_{k=1}^{\infty }g^{k}\epsilon
_{k}^{\left( 3D\right) }.  \tag{26}
\end{equation}

Inserting (24), (25) and (26) into (22) and (23), we obtain a differential
equation taking into account natural units $\left( \hbar =M=\omega =1\right) 
$

\begin{eqnarray}
\epsilon _{k}^{\left( 2D\right) } &=&-\frac{1}{2}\triangledown _{2D}^{2}\phi
_{k}\left( x_{\alpha }\right) +x\ \partial _{x}\phi _{k}\left( x_{\alpha
}\right) +y\ \partial _{y}\phi _{k}\left( x_{\alpha }\right) +igxy^{2}\delta
_{k,1}  \notag \\
&&-\frac{1}{2}\sum_{l=1}^{k-1}\left[ \partial _{x}\phi _{k-l}\left(
x_{\alpha }\right) \partial _{x}\phi _{l}\left( x_{\alpha }\right) +\partial
_{y}\phi _{k-l}\left( x_{\alpha }\right) \partial _{y}\phi _{l}\left(
x_{\alpha }\right) \right] .  \TCItag{27}
\end{eqnarray}%
\begin{eqnarray}
\epsilon _{k}^{\left( 3D\right) } &=&-\frac{1}{2}\triangledown _{3D}^{2}\phi
_{k}\left( x_{\alpha }\right) +x\ \partial _{x}\phi _{k}\left( x_{\alpha
}\right) +y\ \partial _{y}\phi _{k}\left( x_{\alpha }\right) +z\ \partial
_{z}\phi _{k}\left( x_{\alpha }\right)  \notag \\
&&+igxyz\delta _{k,1}-\frac{1}{2}\sum_{l=1}^{k-1}\left[ \partial _{x}\phi
_{k-l}\left( x_{\alpha }\right) \partial _{x}\phi _{l}\left( x_{\alpha
}\right) +\partial _{y}\phi _{k-l}\left( x_{\alpha }\right) \partial
_{y}\phi _{l}\left( x_{\alpha }\right) \right.  \notag \\
&&\left. +\partial _{z}\phi _{k-l}\left( x_{\alpha }\right) \partial
_{z}\phi _{l}\left( x_{\alpha }\right) \right] ,  \TCItag{28}
\end{eqnarray}%
where we have used the abbreviation $\partial _{x_{\alpha }}\equiv \dfrac{%
\partial }{\partial x_{\alpha }}.$

The $\phi _{k}\left( x_{\alpha }\right) $ are expanded in powers of the
coordinates as [3-5,12]%
\begin{equation}
\phi _{k}\left( x,y\right) =\sum_{j=0}^{k}\sum_{m=0}^{k}a_{j,m}^{\left(
k\right) }x^{j}y^{2m}.  \tag{29}
\end{equation}%
\begin{equation}
\phi _{k}\left( x,y,z\right)
=\sum_{j=0}^{k}\sum_{m=0}^{k}\sum_{n=0}^{k}a_{j,m,n}^{\left( k\right)
}x^{j}y^{m}z^{n}.  \tag{30}
\end{equation}%
where $a_{j,m}^{\left( k\right) }$ and $a_{j,m,n}^{\left( k\right) }$ are
non-symmetrical coefficients and can be real \textit{and}/\textit{or}
imaginary.

By inserting (29) and (30) into (27) and (28), we obtain in second order the
following coefficients

\begin{itemize}
\item 2-D
\end{itemize}

For $k=1,$%
\begin{equation}
a_{01}^{\left( 1\right) }=0,\quad a_{10}^{\left( 1\right) }=a_{11}^{\left(
1\right) }=-\frac{i}{3}.  \tag{31}
\end{equation}

For $k=2,$%
\begin{eqnarray}
a_{10}^{\left( 2\right) } &=&a_{11}^{\left( 2\right) }=a_{12}^{\left(
2\right) }=a_{22}^{\left( 2\right) }=0,\ a_{01}^{\left( 2\right) }=-\frac{1}{%
8}  \notag \\
a_{20}^{\left( 2\right) } &=&-\frac{1}{36},\ a_{21}^{\left( 2\right) }=-%
\frac{1}{18},\ a_{02}^{\left( 2\right) }=-\frac{1}{72}.  \TCItag{32}
\end{eqnarray}

\begin{itemize}
\item 3-D
\end{itemize}

For $k=1,$%
\begin{eqnarray}
a_{100}^{\left( 1\right) } &=&a_{010}^{\left( 1\right) }=a_{001}^{\left(
1\right) }=a_{110}^{\left( 1\right) }=a_{101}^{\left( 1\right)
}=a_{011}^{\left( 1\right) }=0,  \notag \\
a_{111}^{\left( 1\right) } &=&-\frac{i}{3}.  \TCItag{33}
\end{eqnarray}

For $k=2,$%
\begin{equation}
a_{200}^{\left( 2\right) }=a_{020}^{\left( 2\right) }=a_{002}^{\left(
2\right) }=a_{220}^{\left( 2\right) }=a_{202}^{\left( 2\right)
}=a_{022}^{\left( 2\right) }=-\frac{1}{72},\mathrm{\ }\text{all others\ }%
a_{j,m,n}^{\left( 2\right) }=0.  \tag{34}
\end{equation}

For $k\geq 3$, we find the recursion relations and the energy correction
coefficients

\begin{itemize}
\item 2-D%
\begin{eqnarray}
a_{j,m}^{\left( k\right) } &=&\frac{1}{2\left( j+m\right) }\left[
a_{j-1,m-1}^{\left( k-1\right) }-2\sum_{k^{\prime }=1}^{k}\left(
a_{2,0}^{\left( k^{\prime }\right) }+a_{0,2}^{\left( k^{\prime }\right)
}\right) a_{j,m}^{\left( k-k^{\prime }\right) }\right.  \notag \\
&&\left. +\left( j+1\right) \left( j+2\right) a_{j+2,m}^{\left( k\right)
}+\left( m+1\right) \left( m+2\right) a_{j,m+2}^{\left( k\right) }\right] , 
\TCItag{35}
\end{eqnarray}%
\begin{equation}
\epsilon _{0}^{\left( 2D\right) }=\frac{1}{2}\left( -1\right) ^{k+1}\left(
a_{2,0}^{\left( 2k\right) }+a_{0,2}^{\left( 2k\right) }\right) .  \tag{36}
\end{equation}

\item 3-D%
\begin{eqnarray}
a_{j,m,n}^{\left( k\right) } &=&\frac{1}{2\left( j+m+n\right) }\left[
a_{j-1,m-1,n-1}^{\left( k-1\right) }-2\sum_{k^{\prime }=1}^{k}\left(
a_{2,0,0}^{\left( k^{\prime }\right) }+a_{0,2,0}^{\left( k^{\prime }\right)
}+a_{0,0,2}^{\left( k^{\prime }\right) }\right) \right.  \notag \\
&&a_{j,m,n}^{\left( k-k^{\prime }\right) }+\left( j+1\right) \left(
j+2\right) a_{j+2,m,n}^{\left( k\right) }+\left( m+1\right) \left(
m+2\right) a_{j,m+2,n}^{\left( k\right) }  \notag \\
&&\left. +\left( n+1\right) \left( n+2\right) a_{j,m,n+2}^{\left( k\right) } 
\right]  \TCItag{37}
\end{eqnarray}%
\begin{equation}
\epsilon _{0}^{\left( 3D\right) }=\frac{1}{2}\left( -1\right) ^{k+1}\left(
a_{2,0,0}^{\left( 2k\right) }+a_{0,2,0}^{\left( 2k\right)
}+a_{0,0,2}^{\left( 2k\right) }\right) .  \tag{38}
\end{equation}
\end{itemize}

Table 1 and Table 2 show the coefficients up to the 10th order [10,12]; one
sees that the results obtained by the Bender-Wu method are in agreement with
those obtained by the connected fourth order Feynman diagrams (see (18) and
(19)).

\bigskip

{\small Table 1}

{\small Weak-coupling coefficients for the 2-dimensional potential up to the
10th order, }$k${\small \ represents the order of expansion and }$\epsilon
_{k}${\small \ coefficients of the energy corrections.}

\begin{center}
\begin{tabular}{|c||c|c|c|c|c|}
\hline
$k$ & $1$ & $2$ & $3$ & $4$ & $5$ \\ \hline
$\epsilon _{k}$ & $0$ & $\frac{5}{24}$ & $0$ & $-\frac{223}{864}$ & $0$ \\ 
\hline\hline
$k$ & $6$ & $7$ & $8$ & $9$ & $10$ \\ \hline
$\epsilon _{k}$ & $\frac{116407}{155520}$ & $0$ & $-\frac{346266143}{%
111974400}$ & $0$ & $\frac{2360833242959}{141087744000}$ \\ \hline
\end{tabular}
\end{center}

\bigskip

{\small Table 2}

{\small Weak-coupling coefficients for the 3-dimensional potential up to the
10th order, }$k${\small \ represents the order of expansion and }$\epsilon
_{k}${\small \ coefficients of the energy corrections.}

\begin{center}
$%
\begin{tabular}{|c||c|c|c|c|c|}
\hline
$k$ & $1$ & $2$ & $3$ & $4$ & $5$ \\ \hline
$\epsilon _{k}$ & $0$ & $\frac{1}{24}$ & $0$ & $-\frac{5}{576}$ & $0$ \\ 
\hline\hline
$k$ & $6$ & $7$ & $8$ & $9$ & $10$ \\ \hline
$\epsilon _{k}$ & $\frac{5069}{622080}$ & $0$ & $-\frac{2441189}{289598400}$
& $0$ & $\frac{8034211571}{752467968000}$ \\ \hline
\end{tabular}%
$

\bigskip
\end{center}

\section{Resummation procedure : Strong-coupling limit}

In this section, we are interested to resume the perturbation series (18)
and (19) to the strong-coupling limit. To this end, substituting the
coordinate $x_{\alpha },$ where $\alpha =1,2,3,$\ by%
\begin{equation}
x_{\alpha }\longrightarrow g^{-1/5}x_{\alpha },  \tag{39}
\end{equation}%
the Schr\"{o}dinger equations (20) and (21) become%
\begin{equation}
-\frac{\hbar ^{2}}{2M}\nabla _{2D}^{2}\psi \left( x,y\right) +\left[ \frac{%
M\ \omega ^{2}}{2}g^{-4/5}\left( x^{2}+y^{2}\right) +ixy^{2}\right] \psi
\left( x,y\right) =g^{-4/5}E_{2D}\psi \left( x,y\right) .  \tag{40}
\end{equation}%
\begin{gather}
-\frac{\hbar ^{2}}{2M}\nabla _{3D}^{2}\psi \left( x,y,z\right) +\left[ \frac{%
M\ \omega ^{2}}{2}g^{-4/5}\left( x^{2}+y^{2}+z^{2}\right) +ixyz\right] \psi
\left( x,y,z\right)  \notag \\
=g^{-4/5}E_{3D}\psi \left( x,y,z\right) .  \tag{41}
\end{gather}

The wave functions and the energy will be expanded in powers of the coupling
constant $g$ and yield [13,15]

\begin{equation}
\psi \left( x_{\alpha }\right) =\psi _{0}\left( x_{\alpha }\right)
+g^{-4/5}\psi _{1}\left( x_{\alpha }\right) +g^{-8/5}\psi _{2}\left(
x_{\alpha }\right) +\cdots .  \tag{42}
\end{equation}%
\begin{equation}
E_{D}=g^{2/5}\left( \epsilon _{0}+g^{-4/5}\epsilon _{1}+g^{-8/5}\epsilon
_{2}+\cdots \right) .  \tag{43}
\end{equation}

The strong-coupling coefficients of the ground-state for the potentials (2)
and (3) can be obtained by resuming the weak-coupling series [10,12]
obtained in the last two sections. Up to the fourth order, the weak-coupling
series read%
\begin{equation}
\ E_{2D}^{\left( 0\right) }=\hbar \omega +\frac{5}{24}\frac{g^{2}\hbar ^{2}}{%
M^{3}\omega ^{4}}-\frac{223}{864}\frac{g^{4}\hbar ^{3}}{M^{6}\omega ^{9}}%
+O\left( g^{6}\right) .  \tag{44}
\end{equation}%
\begin{equation}
E_{3D}^{\left( 0\right) }=\frac{3}{2}\hbar \omega +\frac{1}{24}\frac{%
g^{2}\hbar ^{2}}{M^{3}\omega ^{4}}-\frac{7}{576}\frac{g^{4}\hbar ^{3}}{%
M^{6}\omega ^{9}}+O\left( g^{6}\right) .  \tag{45}
\end{equation}

The power behavior of $E_{D}^{\left( 0\right) }$\ taken in (43) is
independent of the order $N$\ and solely the coefficients depend on $N$%
\begin{equation}
E_{D}^{\left( N\right) }=g^{2/5}\left( \epsilon _{0}^{\left( N\right)
}+g^{-4/5}\epsilon _{1}^{\left( N\right) }+g^{-8/5}\epsilon _{2}^{\left(
N\right) }+\cdots \right) .  \tag{46}
\end{equation}

In order to perform a resummation, an artificial parameter trick is
introduced often called Kleinert's square-root [2,13-15]%
\begin{equation}
\omega \rightarrow \Omega \sqrt{1+g^{2}r},  \tag{47}
\end{equation}%
with%
\begin{equation}
r=\frac{\omega ^{2}-\Omega ^{2}}{g^{2}\Omega ^{2}}.  \tag{48}
\end{equation}

\ Therefore, inserting (47) into (44) and (45) and re-expanding in coupling
constant $g^{2}$ to the first order taking into account (48), we obtain%
\begin{equation}
E_{2D}^{\left( 1\right) }\left( g,\omega ,\Omega \right) =\frac{\hbar \Omega 
}{2}+\frac{\hbar \omega ^{2}}{2\Omega }+g^{2}\frac{5\hbar ^{2}}{24\Omega ^{4}%
}+O\left( g^{4}\right) .  \tag{49}
\end{equation}%
\begin{equation}
E_{3D}^{\left( 1\right) }\left( g,\omega ,\Omega \right) =\frac{3\hbar
\Omega }{2}+\frac{3\hbar \omega ^{2}}{4\Omega }+g^{2}\frac{\hbar ^{2}}{%
24\Omega ^{4}}+O\left( g^{4}\right) .  \tag{50}
\end{equation}

Performing the first derivative with respect to the variational parameter $%
\Omega $ and considering its strong-coupling behavior [13,15]%
\begin{equation}
\Omega =g^{2/5}\left( \Omega _{0}+g^{-4/5}\Omega _{1}+g^{-8/5}\Omega
_{2}+\cdots \right)  \tag{51}
\end{equation}%
we get

\begin{itemize}
\item 2-D%
\begin{equation}
\Omega _{0}=\sqrt[5]{\frac{5\hbar }{3}}\quad ;\quad \Omega _{1}=\frac{\omega
^{2}}{5}\sqrt[5]{\frac{3}{5\hbar }}\quad ;\quad \Omega _{2}=\frac{\omega ^{4}%
}{25}\sqrt[5]{\frac{27}{125\hbar ^{3}}}.  \tag{52}
\end{equation}

\item 3-D%
\begin{equation}
\Omega _{0}=\sqrt[5]{\frac{2\hbar }{9}}\quad ;\quad \Omega _{1}=\frac{\omega
^{2}}{5}\sqrt[5]{\frac{9}{2\hbar }}\quad ;\quad \Omega _{2}=\frac{\omega ^{4}%
}{25}\sqrt[5]{\frac{3}{8\hbar ^{3}}}.  \tag{53}
\end{equation}
\end{itemize}

Inserting these results into (49) and (50) yield the strong-coupling
coefficients of the ground-state energy $\epsilon _{k}^{\left( N\right) }$

\begin{itemize}
\item 2-D%
\begin{equation}
\epsilon _{0}^{\left( 1\right) }=\frac{5}{8}\sqrt[5]{\frac{5\hbar ^{6}}{3}}%
\approx 0.69222\hbar ^{6/5};\quad \epsilon _{1}^{\left( 1\right) }=\frac{%
\omega ^{2}}{2}\sqrt[5]{\frac{3\hbar ^{4}}{5}};\quad \epsilon _{2}^{\left(
1\right) }=\frac{\omega ^{4}}{20}\sqrt[5]{\frac{3\hbar ^{2}}{5}}.  \tag{54}
\end{equation}

\item 3-D%
\begin{equation}
\epsilon _{0}^{\left( 1\right) }=\frac{5}{8}\sqrt[5]{\frac{27\hbar ^{6}}{16}}%
\approx 0.69395\hbar ^{6/5};\quad \epsilon _{1}^{\left( 1\right) }=\frac{%
3\omega ^{2}}{4}\sqrt[5]{\frac{9\hbar ^{4}}{2}};\quad \epsilon _{2}^{\left(
1\right) }=-\frac{9\omega ^{4}}{40}\sqrt[5]{\frac{3\hbar ^{2}}{8}}.  \tag{55}
\end{equation}
\end{itemize}

However, it turns out that the numerical values of the leading
strong-coupling coefficients obtained in (54) and (55) are lower than those
calculated in weak-coupling limit (see (18) and (19)). Thus, the convergence
of the VPT-results is less satisfactory. Following [13], we must rely upon
that the farfetched values must be higher than those calculated in
weak-coupling limit. It is what we are going to show in section 5.

\section{D-dimensional complex effective potential}

For any given potential, the corresponding effective potential is obtained
by performing a Legendre transformation of the ground-state energy in the
special case of the external current being constant [2,15]; it will be
expanded in powers of $\hbar $ rather than the coupling constant $g$ and
depends on the new parameter $X_{\alpha },\ $where $\alpha =1,2,3,$ namely
the path average,%
\begin{eqnarray}
V_{\text{eff}}\left( X_{\alpha }\right) &\equiv &\sum_{l=0}^{\infty }\hbar
^{l}V^{\left( l\right) }\left( X_{\alpha }\right)  \notag \\
&=&V\left( X_{\alpha }\right) +\frac{\hbar }{2}\sum_{\alpha }\func{tr}\ln
G_{X_{\alpha }}^{-1}+V_{D}^{\left( \text{int}\right) }\left( X_{\alpha
}\right) .  \TCItag{56}
\end{eqnarray}%
where the superscript $l$ indicates the number of loops involved in Feynman
diagrams, the trace-logarithm functions are given by the ground-state energy
of harmonic oscillators, they are connected to the partial frequencies by%
\begin{eqnarray}
\frac{\hbar }{2}\func{tr}\ln G_{X_{\alpha }}^{-1} &\equiv &\frac{\hbar 
\widetilde{\omega }_{X_{\alpha }}}{2}  \notag \\
&=&\frac{\hbar }{2}\sqrt{\partial _{X_{\alpha }}^{2}V_{D}\left( X_{\alpha
}\right) }.  \TCItag{57}
\end{eqnarray}

The frequency of the propagator is given as the sum of all partial
frequencies%
\begin{equation}
\widetilde{\omega }_{D}=\sqrt{\sum_{\alpha }\widetilde{\omega }_{X_{\alpha
}}^{2}}.  \tag{58}
\end{equation}

$V^{\left( \text{int}\right) }\left( X_{\alpha }\right) $ is called
interaction potential [2,13-15] and contains all one-particle irreducible
vacuum diagrams.

\subsection{Weak-coupling limit}

The aim of this section is to deduce weak-coupling ground-state energy for
the potentials (2) and (3) by using the effective potential [10,12]. The
computation will be performed until the third-loop order at low
temperatures, i.e. $\left( T\rightarrow 0\right) .$

The frequency of the propagator of the potentials (2) and (3) are now given,
using (57) and (58), by%
\begin{equation}
\widetilde{\omega }_{2D}=\sqrt{2\omega ^{2}+2igX}.  \tag{59}
\end{equation}%
\begin{equation}
\widetilde{\omega }_{3D}=\sqrt{3}\omega .  \tag{60}
\end{equation}

The corresponding interaction potentials are read%
\begin{equation}
V_{2D}^{\left( \text{int}\right) }\left( X,Y\right) =-\underset{T\rightarrow
0}{\lim }k_{B}T\ \left[ \frac{1}{36}\FRAME{itbpF}{0.6129cm}{0.6129cm}{%
0.2109cm}{}{}{f_2.jpg}{\special{language "Scientific Word";type
"GRAPHIC";maintain-aspect-ratio TRUE;display "PICT";valid_file "F";width
0.6129cm;height 0.6129cm;depth 0.2109cm;original-width
0.2084in;original-height 0.2084in;cropleft "0";croptop "1.0200";cropright
"1.0200";cropbottom "0";filename 'Feyndiag/f_2.JPG';file-properties "XNPEU";}%
}+\frac{5}{1296}\FRAME{itbpF}{0.5865cm}{0.6129cm}{0.2109cm}{}{}{f_4.jpg}{%
\special{language "Scientific Word";type "GRAPHIC";maintain-aspect-ratio
TRUE;display "PICT";valid_file "F";width 0.5865cm;height 0.6129cm;depth
0.2109cm;original-width 0.198in;original-height 0.2084in;cropleft
"0";croptop "1.0200";cropright "1.0210";cropbottom "0";filename
'Feyndiag/f_4.JPG';file-properties "XNPEU";}}+\frac{1}{648}\FRAME{itbpF}{%
0.5865cm}{0.6129cm}{0.2109cm}{}{}{f_5.jpg}{\special{language "Scientific
Word";type "GRAPHIC";maintain-aspect-ratio TRUE;display "PICT";valid_file
"F";width 0.5865cm;height 0.6129cm;depth 0.2109cm;original-width
0.198in;original-height 0.2084in;cropleft "0";croptop "1.0200";cropright
"1.0210";cropbottom "0";filename 'Feyndiag/f_5.JPG';file-properties "XNPEU";}%
}\right] +O\left( \hbar ^{4}\right) .  \tag{61}
\end{equation}%
\begin{equation}
V_{3D}^{\left( \text{int}\right) }\left( X,Y,Z\right) =-\underset{%
T\rightarrow 0}{\lim }k_{B}T\ \left[ \frac{1}{72}\FRAME{itbpF}{0.6129cm}{%
0.6129cm}{0.2109cm}{}{}{f_2.jpg}{\special{language "Scientific Word";type
"GRAPHIC";maintain-aspect-ratio TRUE;display "PICT";valid_file "F";width
0.6129cm;height 0.6129cm;depth 0.2109cm;original-width
0.2084in;original-height 0.2084in;cropleft "0";croptop "1.0200";cropright
"1.0200";cropbottom "0";filename 'Feyndiag/f_2.JPG';file-properties "XNPEU";}%
}+\frac{1}{1728}\FRAME{itbpF}{0.5865cm}{0.6129cm}{0.2109cm}{}{}{f_4.jpg}{%
\special{language "Scientific Word";type "GRAPHIC";maintain-aspect-ratio
TRUE;display "PICT";valid_file "F";width 0.5865cm;height 0.6129cm;depth
0.2109cm;original-width 0.198in;original-height 0.2084in;cropleft
"0";croptop "1.0200";cropright "1.0210";cropbottom "0";filename
'Feyndiag/f_4.JPG';file-properties "XNPEU";}}+\frac{1}{5184}\FRAME{itbpF}{%
0.5865cm}{0.6129cm}{0.2109cm}{}{}{f_5.jpg}{\special{language "Scientific
Word";type "GRAPHIC";maintain-aspect-ratio TRUE;display "PICT";valid_file
"F";width 0.5865cm;height 0.6129cm;depth 0.2109cm;original-width
0.198in;original-height 0.2084in;cropleft "0";croptop "1.0200";cropright
"1.0210";cropbottom "0";filename 'Feyndiag/f_5.JPG';file-properties "XNPEU";}%
}\right] +O\left( \hbar ^{4}\right) .  \tag{62}
\end{equation}%
where the previous diagrams are deduced from new-Feynman laws

\begin{itemize}
\item for propagator%
\begin{equation}
\FRAME{itbpF}{2.4229cm}{0.9951cm}{0.5008cm}{}{}{f_7.jpg}{\special{language
"Scientific Word";type "GRAPHIC";maintain-aspect-ratio TRUE;display
"PICT";valid_file "F";width 2.4229cm;height 0.9951cm;depth
0.5008cm;original-width 0.9167in;original-height 0.3649in;cropleft
"0";croptop "0.9942";cropright "1.0045";cropbottom "0";filename
'Feyndiag/f_7.JPG';file-properties "XNPEU";}}\rightarrow \quad G_{\omega
}\left( \tau _{a},\tau _{b}\right) .  \tag{63}
\end{equation}

\item for vertices%
\begin{equation}
\FRAME{itbpF}{1.8364cm}{1.81cm}{0.905cm}{}{}{f_8.jpg}{\special{language
"Scientific Word";type "GRAPHIC";maintain-aspect-ratio TRUE;display
"PICT";valid_file "F";width 1.8364cm;height 1.81cm;depth
0.905cm;original-width 0.6875in;original-height 0.6772in;cropleft
"0";croptop "1.0061";cropright "1.0060";cropbottom "0";filename
'Feyndiag/f_8.JPG';file-properties "XNPEU";}}\rightarrow \quad \frac{-D}{%
\hbar }\sum_{\alpha =i}^{D}\sum_{\alpha =j}^{D}\sum_{\alpha =k}^{D}\partial
_{i}\partial _{j}\partial _{k}\ V\left( X_{\alpha }\right) .  \tag{64}
\end{equation}
\end{itemize}

Substituting (59)-(62) into (56) and taking into account (63) and (64), we
obtain the expressions of effective potential at low temperatures%
\begin{multline}
\underset{T\rightarrow 0}{\lim }V_{\text{eff}}^{\left( \text{2D}\right)
}\left( X,Y\right) =\frac{\omega ^{2}}{2}\left( X^{2}+Y^{2}\right) +igXY^{2}+%
\frac{\hbar }{2}\omega +\frac{\hbar }{2}\sqrt{\omega ^{2}+2igX}+  \notag \\
\frac{\hbar ^{2}g^{2}}{3\left( 2\omega ^{2}+2igX\right) ^{3/2}}-\left[ \frac{%
2}{3}\frac{1}{648}+\frac{22}{27}\frac{5}{1296}\right] \frac{324\sqrt{2}\hbar
^{3}g^{4}}{\left( 2\omega ^{2}+2igX\right) ^{9/2}}+O\left( \hbar ^{4}\right)
.  \tag{65}
\end{multline}%
\begin{eqnarray}
\underset{T\rightarrow 0}{\lim }V_{\text{eff}}^{\left( \text{3D}\right)
}\left( X,Y,Z\right) &=&\frac{\omega ^{2}}{2}\left( X^{2}+Y^{2}+Z^{2}\right)
+igXYZ+\frac{3\hbar \omega }{2}+\frac{\hbar ^{2}g^{2}}{24\omega ^{4}}  \notag
\\
&&-\left[ \frac{2}{3}\frac{1}{5184}+\frac{22}{27}\frac{5}{1728}\right] \frac{%
81\hbar ^{3}g^{4}}{4\omega ^{9}}+O\left( \hbar ^{4}\right) .  \TCItag{66}
\end{eqnarray}

The path average $X_{\alpha }$ in the case of complex potentials can be
expanded in the form [13,15]%
\begin{equation}
X_{\alpha }=i\ \left( X_{\alpha 0}+\hbar X_{\alpha 1}+\hbar ^{2}X_{\alpha
2}+\cdots \right) .  \tag{67}
\end{equation}

Inserting this identity into (65) and (66), and performing the first
derivative with respect to $X_{\alpha \text{ }}$and setting the resulting
expressions to zero, we get

\begin{itemize}
\item 2-D%
\begin{eqnarray}
X_{10} &\equiv &X_{0}=0;\quad X_{11}\equiv X_{1}=-\frac{g}{2\omega ^{3}}%
;\quad X_{12}\equiv X_{2}=\frac{5g^{3}}{12\omega ^{3}}.  \TCItag{68.a} \\
X_{20} &\equiv &Y_{0}=0;\quad X_{21}\equiv Y_{1}=0;\quad \qquad \ \
X_{22}\equiv Y_{2}=0.  \TCItag{68.b}
\end{eqnarray}

\item 3-D%
\begin{equation}
X_{\alpha k}\equiv 0.  \tag{69}
\end{equation}

for $\alpha =1,2,3$, and $k=0,1,2$.
\end{itemize}

Re-inserting (68) and (69), respectively, into (65) and (66), and expanding
until the third--order in $\hbar $ yields the ground-state energy at low
temperatures%
\begin{equation}
E_{2D}^{\left( 0\right) }=\hbar \omega +\frac{5}{24}\frac{g^{2}\hbar ^{2}}{%
M^{3}\omega ^{4}}-\frac{223}{864}\frac{g^{4}\hbar ^{3}}{M^{6}\omega ^{9}}%
+O\left( g^{6}\right) .  \tag{70}
\end{equation}%
\begin{equation}
E_{3D}^{\left( 0\right) }=\frac{3}{2}\hbar \omega +\frac{1}{24}\frac{%
g^{2}\hbar ^{2}}{M^{3}\omega ^{4}}-\frac{7}{576}\frac{g^{4}\hbar ^{3}}{%
M^{6}\omega ^{9}}+O\left( g^{6}\right) .  \tag{71}
\end{equation}

The results are in agreement with those obtained in sections 2 and 3.

\subsection{Strong-coupling limit}

Let us now resume the effective potential. Since the effective potential is
expanded in powers of $\hbar $, Kleinert's square-root will be converted
accordingly to [13-15]%
\begin{equation}
\omega \rightarrow \Omega \sqrt{1+\hbar \ r}.  \tag{72}
\end{equation}%
with%
\begin{equation}
r=\frac{\omega ^{2}-\Omega ^{2}}{\hbar \ \Omega ^{2}}.  \tag{73}
\end{equation}

\subsubsection{Two-dimensional case}

\paragraph{First-order}

Substituting (72) and (73) into (65) at first order in $\hbar $, we obtain%
\begin{equation}
V_{\text{eff},1}^{\left( 2D\right) }\left( X,Y\right) =\frac{\omega ^{2}}{2}%
\left( X^{2}+Y^{2}\right) +igXY^{2}+\frac{\hbar \Omega }{2}+\frac{\hbar }{2}%
\sqrt{\Omega ^{2}+2igX}.  \tag{74}
\end{equation}

We now optimize in $\Omega $ and $X_{\alpha }$; the resulting equations
allow us to determine the strong-coupling behavior of $X_{\alpha }$ given by
[13,15]%
\begin{equation}
X_{\alpha }=-ig^{-1/5}\left( X_{\alpha 0}+X_{\alpha 1}g^{-4/5}+X_{\alpha
2}g^{-8/5}+\cdots \right) .  \tag{75}
\end{equation}%
where the corresponding coefficients read as%
\begin{eqnarray}
X_{0} &=&0.417288\ \hbar ^{6/5};\ X_{1}=-0.245404\ \omega ^{2};\
X_{2}=0.063624\ \omega ^{2}\hbar ^{-6/5}.  \TCItag{76.a} \\
Y_{k} &=&\sqrt{2}\ X_{k};\quad \qquad k=0,1,2.  \TCItag{76.b}
\end{eqnarray}

Re-inserting the results (76) into (75), and again in (74), we obtain the
strong-coupling behavior of the ground-state energy%
\begin{equation}
\epsilon _{0}^{\left( 1\right) }\approx 1.1263168\ \hbar ^{6/5}.  \tag{77}
\end{equation}

\paragraph{Second order}

In second order, the effective potential becomes%
\begin{eqnarray}
V_{\text{eff},2}^{\left( 2D\right) }\left( X,Y\right) &=&\frac{\omega ^{2}}{2%
}\left( X^{2}+Y^{2}\right) +igXY^{2}+\frac{\hbar }{4}\left[ \frac{\omega ^{2}%
}{\Omega ^{2}}+\Omega +\frac{\omega ^{2}+\Omega ^{2}+4igX}{\sqrt{\Omega
^{2}+2igX}}\right]  \notag \\
&&+\frac{g^{2}\hbar ^{2}}{12\left( \Omega ^{2}+igX\right) ^{2}}.  \TCItag{78}
\end{eqnarray}

Following the same steps as was done in the first order, the strong-coupling
behavior of the ground-state energy in second order is%
\begin{equation}
\epsilon _{0}^{\left( 2\right) }\approx 1.13595605\ \hbar ^{6/5}.  \tag{79}
\end{equation}

\subsubsection{Three-dimensional case}

\paragraph{First order}

Substituting (72) and (73) into (66) at first order in $\hbar $, we obtain
the effective potential

\begin{equation}
V_{\text{eff},1}^{\left( 3D\right) }\left( x\right) =\frac{\omega ^{2}}{2}%
\left( X^{2}+Y^{2}+Z^{2}\right) +igXYZ+\frac{3\hbar \Omega }{2}.  \tag{80}
\end{equation}

Let us now optimize in $\Omega $ and $X_{\alpha }$, with $\alpha =1,2,3$. We
can note that, from (80), the first derivative with respect to $\Omega $ is
constant $\left( =\dfrac{3\hbar }{2}\right) $ and different than zero. In
order to avoid this ambiguity, we need to substitute $\Omega $ in (80) by a
mathematical trick according to%
\begin{equation}
\Omega \rightarrow \sqrt{\Omega ^{2}+2\ ig\lambda X},  \tag{81}
\end{equation}%
where $\lambda \ll 1$ and by performing $\lambda $-expansion around zero $%
\left( \lambda \rightarrow 0\right) $, we obtain%
\begin{eqnarray}
X_{0} &=&\frac{1}{\lambda }\sqrt[5]{\frac{\hbar ^{2}}{648}};\quad \quad
X_{1}=\frac{\omega ^{2}}{5\lambda };\quad \ X_{2}=\frac{\omega ^{2}\left( 20+%
\sqrt[5]{648}\omega ^{2}\right) }{100\lambda }\hbar ^{-6/5}.  \TCItag{82.a}
\\
Y_{0} &=&0;\qquad \qquad \quad \ Y_{1}=\omega ^{2};\qquad Y_{2}=\hbar
^{-6/5}.  \TCItag{82.b} \\
Z_{0} &=&-\frac{1}{\lambda }\sqrt[5]{\frac{\hbar ^{2}}{648}};\quad Z_{1}=-%
\frac{\omega ^{2}}{5\lambda };\quad \ \ Z_{2}=-\frac{3\omega ^{4}}{50\lambda 
}\sqrt[5]{\frac{81}{5\hbar ^{6}}}.  \TCItag{82.c}
\end{eqnarray}

Re-inserting (81) and (82) into (80), taking into account (75), we obtain
the strong-coupling behavior of the ground-state energy at first-order%
\begin{equation}
\epsilon _{0}^{\left( 1\right) }=\frac{3^{3/5}}{2^{3/10}}\ \hbar
^{6/5}\approx 1.5702317\ \hbar ^{6/5}.  \tag{83}
\end{equation}

\paragraph{Second order}

Substituting (72) and (73) into (66), and expanding the result\ until the
second--order in $\hbar $\ yields the effective potential%
\begin{equation}
V_{\text{eff},2}^{\left( 3D\right) }\left( X,Y,Z\right) =\frac{\omega ^{2}}{2%
}\left( X^{2}+Y^{2}+Z^{2}\right) +igXYZ+\frac{3\left( \omega ^{2}+\Omega
^{2}\right) \hbar }{4\Omega }+\frac{\hbar ^{2}}{24\Omega ^{4}}.  \tag{84}
\end{equation}

Again, the strong-coupling behavior of the ground-state energy in second
order is obtainable following the same steps as was done in the first order,
we obtain%
\begin{equation}
\epsilon _{0}^{\left( 2\right) }=\frac{3^{3/5}}{2^{3/10}}\frac{\sqrt{1+\sqrt{%
230}}}{4}\ \hbar ^{6/5}\approx 1.5783441\ \hbar ^{6/5}.  \tag{85}
\end{equation}

In both cases, one can sees that the new values of the leading
strong-coupling coefficients are in good agreement with those expected with
a small deviation from the first order, i.e. $\dfrac{\left\vert \epsilon
_{0}^{\left( 2\right) }-\epsilon _{0}^{\left( 1\right) }\right\vert }{%
\epsilon _{0}^{\left( 1\right) }}$, in order of $0.8\%$ and $0.5\%$,
respectively.

\section{Conclusion}

In sections 2 and 3, we have derived the weak-coupling coefficients for the
ground-state energy for the potentials (2) and (3) using the Feynman
diagrammatical expansion and Bender-Wu recursion relations. Both results are
in agreement with those obtained by the old-fashion perturbation theory,
namely the Rayleigh-Schr\"{o}dinger method. We have proceeded in section 4
to resume the weak-coupling series (70) and (71) to the strong-coupling
limit by applying Kleinert's square-root trick. However, the leading
strong-coupling coefficients for the ground-state energy lie below those
calculated in weak-coupling limit, thus, the rate of convergence is less
satisfactory. We introduce in section 5 the effective potential which, once
combined with the VPT, allows to determine the same weak-coupling series
obtained in sections 2 and 3. In order to recover the rate of convergence in
the strong-coupling limit, we proceed to resume the effective potential. It
turns out that VPT-results for leading strong-coupling coefficients obtained
in second order (79) and (85) approaches those expected with a very small
deviation compared to the first order (77) and (83), thanks to the
introduction of the path average variational parameter.

Following [16], it is interesting to apply the covariant effective potential
for particle moving in one-dimensional complex cubic potential for solving
the corresponding Schr\"{o}dinger equation with position-dependant mass.


\begin{thebibliography}{10}
\bibitem[1]{} R. P. Feynman, A. R. Hibbs, Quantum Mechanics and Path
Integral, McGraw, New-York, 1965.

\bibitem[2]{} H. Kleinert, Path Integrals in Quantum Mechanics, Statistics,
Polymers Physics and Financial Markets, Fouth Ed., World Scientific,
Singapore, 2006.

\bibitem[3]{} C. M. Bender, T. T. Wu, Phys. Rev. 184 (1969) 1231.

\bibitem[4]{} C. M. Bender, T. T. Wu, Phys. Rev. Lett. 27 (1971) 461.

\bibitem[5]{} C. M. Bender, T. T. Wu, Phys. Rev. D7 (1973) 1620.

\bibitem[6]{} R. P. Feynman, H. Kleinert, Phys. Rev. A 34 (1986) 5080.

\bibitem[7]{} H. Kleinert, Phys. Lett. B 280 (1992) 251.

\bibitem[8]{} H. Kleinert, Phys. Lett. A 173 (1993) 332.

\bibitem[9]{} M. Bachmann, H. Kleinert, A. Pelster, Phys. Rev. A 60 (1999)
3429.

\bibitem[10]{} M. Bentaiba, S.-A. Yahiaoui, L. Chetouani, Phys. Lett. A 331
(2004) 175.

\bibitem[11]{} M. Bentaiba, L. Chetouani, A. Mazouz, Phys. Lett. A 295
(2002) 13.

\bibitem[12]{} C. M. Bender, G. V. Dunne, P. N. Meisinger, M. Simsek, Phys.
Lett. A 281 (2001) 311.

\bibitem[13]{} S. F. Brandt, H. Kleinert, A. Pelster, J. Math. Phys. 46
(2005) 032101.

\bibitem[14]{} S. F. Brandt, A. Pelster, J. Math. Phys. 46 (2005) 112105.

\bibitem[15]{} S. F. Brandt, Diploma thesis, Fachbereich Physik, Freie
Universit\"{a}t, Berlin (May 2004).

\bibitem[16]{} H. Kleinert, A. Chervyakov, Phys. Lett. A 299 (2002) 319.
\end{thebibliography}
\end{document}